\documentstyle[aps,twocolumn,epsfig,eqsecnum]{revtex}

\begin{document}
\draft

\wideabs{

  \title{Entanglement by a beam splitter: nonclassicality as a prerequisite for entanglement
}

  \author{M. S. Kim,$^1$ W. Son$^{1,2}\footnote{Also at Institute of Quantum Information Processing and Systems,  
    The University of Seoul, \\
    Dongdaemun-Gu, Seoul, Korea}, $ V. Bu\v zek,$^{3}$\footnote{Also at Faculty of Informatics, Masaryk University, Botanick\'a 68a,
    602 00 Brno, Czech Republic}, P. L. Knight,$^{4}$
        }

  \address{$^1$ School of Mathematics and Physics, The Queen's University,
    Belfast BTU7 1NNE, United Kingdom \\
    $^2$ Department of Physics, Sogang University, CPO Box 1142,
    Seoul, Korea \\
    $^3$ Institute of Physics, Slovak Academy of Sciences, D\'ubravsk\'a cesta 9,
    842 28 Bratislava, Slovakia \\
    $^4$ Optics Section, Blackett Laboratory, Imperial College, London SW7 2BW, United Kingdom
    }

  \date{\today}

  \maketitle

\begin{abstract}
    A beam splitter is a simple, readily available device which can
    act to entangle the output optical fields.
We show that a necessary condition for the fields at the output
of the beam splitter to be entangled is that the pure input states
exhibit nonclassical behavior.   We generalize this proof for
arbitrary (pure or impure) Gaussian input states. Specifically,
nonclassicality of the input Gaussian fields is a necessary
condition for entanglement of the field modes with the help of
the beam splitter. We conjecture that this is a general property
of the beam splitter: Nonclassicality of the inputs is a necessary
condition for entangling fields in the beam splitter.
\end{abstract}

}


\section{INTRODUCTION}
Entanglement is at the heart of current development of quantum
information processing \cite{BennettDiVincenzo}.
Entanglement-assisted communication can enlarge the channel
capacity \cite{BennettWiesner} and enhance channel efficiency
\cite{BennettBrassard}.  Entanglement may play a key role in
secure communication \cite{Ekert}.  In quantum computation, of
course, qubits are massively entangled.

The generation and characterization of entanglement has been
studied extensively. In particular, a recent experimental advance
realized the generation and distillation of
polarization-entangled photons toward optimal entanglement in a
$2\times 2$-dimensional Hilbert space \cite{Kwiat01}.  The
polarization-entangled photons are generated using type I or type II
parametric down conversion.  The parametric down conversion is
also a standard technique to produce a two-mode squeezed state,
which is an entangled state in an infinite dimensional Hilbert
space \cite{Furu}.

The beam splitter is also one of only a few
experimentally-accessible devices which may act as an entangler.
There have been some previous studies of a beam splitter as an
entangler \cite{Tan,Sanders,Paris}.  In particular, Paris
\cite{Paris} studied entanglement properties of the output state
from a Mach-Zehnder interferometer for squeezed input states. The
action of a linear directional coupler can also be described by
the beam splitter operator.  Photon statistics and nonclassical
properties of the output fields from a linear directional coupler
were studied for Fock and squeezed inputs \cite{Lai}.

In this paper we investigate the entangling properties of a beam
splitter for various pure input states including Fock states and
squeezed states.  We find a simple formula to determine the
entanglement of output fields for squeezed input fields. We also
study the entanglement of output fields when the input fields are
in a Gaussian mixed state and provide a sufficient condition for
input fields to have no entanglement in the output state: when
two Gaussian ``classical'' fields are input to the beam splitter,
the output state is never entangled. We find that entanglement of
the output state is strongly related to the nonclassicality of
the input fields.

\section{Fock-state input}
Fig. 1 shows the schematic arrangement of a beam splitter.  The
input field described by the operator $\hat{a}$ is superposed on
the other input field with operator $\hat{b}$ by a lossless
symmetric beam splitter, with amplitude reflection and
transmission coefficients $r$ and $t$. The output-field
annihilation operators are given by
\begin{equation}
\label{eq:bs-mode}
\hat{c}=\hat{B}\hat{a}\hat{B}^\dag~~~,~~~\hat{d}=\hat{B}\hat{b}\hat{B}^\dag
\end{equation}
where the beam splitter operator is \cite{Campos89}
\begin{equation}
\label{eq:Fock-beamsplitter}
\hat{B}=\exp\left[{\theta \over 2}(\hat{a}^\dag\hat{b}\mbox{e}^{i\phi}-\hat{a}\hat{b}^\dag\mbox{e}^{-i\phi})\right]
\end{equation}
with the amplitude reflection and transmission coefficients
\begin{equation}
\label{eq:t-r}
t=\cos{\theta \over 2}~~~,~~~r=\sin{\theta \over 2}.
\end{equation}
The beam splitter gives the phase difference $\phi$ between the reflected
and transmitted fields.

In this paper we are interested in entanglement properties of the
output state.  Suppose that input states are two independent Fock
states, $|n_1,n_2\rangle\equiv|n_1\rangle_a|n_2\rangle_b$. The
output fields are then a superposition of two-mode Fock states:
\begin{eqnarray}
\label{eq:Fock-output}
\hat{B}|n_1,n_2\rangle &=& \sum_{N_1~N_2}|N_1,N_2\rangle\langle N_1, N_2|\hat{B}|n_1, n_2\rangle
\nonumber \\
&=&\sum_{N_1~N_2}B_{n_1n_2}^{N_1N_2}
|N_1, N_2\rangle
\end{eqnarray}
where
\begin{eqnarray}
\label{eq:Fock-output-1}
B_{n_1n_2}^{N_1N_2} &=& \mbox{e}^{-i\phi(n_1-N_1)}\sum_{k=0}^{n_1}\sum_{l=0}^{n_2}(-1)^{n_1-k}r^{n_1+n_2-k-l}t^{k+l}
\nonumber \\
\times&&\frac{\sqrt{n_1!n_2!N_1!N_2!}}{k!(n_1-k)!l!(n_2-l)!}\delta_{N_1,n_2+k-l}\delta_{N_2,n_1-k+l}
\end{eqnarray}
with $\delta$ a Kronecker delta function.
When the total number of input photons is $N=n_1+n_2$, the output state becomes an ($N+1$)-dimensional entangled state.

The von Neumann entropy is a measure of entanglement for pure
bipartite states (See e.g. \cite{PhoenixKnight}), which becomes
$\ln(N+1)$ when an ($N+1$)-dimensional bipartite system is maximally
entangled. The von Neumann entropy $E(\hat{\rho}_a)$ for the
reduced density operator
$\hat{\rho}_a=\mbox{Tr}_b\hat{B}|n_1,n_2\rangle\langle
n_1,n_2|B^\dag$ is
\begin{equation}
\label{eq:von-Neumann}
E(\hat{\rho}_a)=-\sum_{N_1~N_2}\left| B_{n_1n_2}^{N_1N_2}\right|^2\ln \left| B_{n_1n_2}^{N_1N_2}\right|^2.
\end{equation}

Fig. 2 shows the von Neumann entropy $E(\hat{\rho}_a)$ is a
function of the reflection coefficient $r$ and configuration of
input photon numbers.  It is interesting to note that the entropy
does not necessarily maximize for a 50:50 beam splitter.  This is
discussed further in the following subsections.

\subsection{SU(2) coherent state}
When $N$ number of photons are injected into one input port while no photon is injected into the other input port,
the output state turns into a state generally known as an SU(2) coherent state \cite{Wodkiewicz,Buzek}.
Substituting $n_1=0$ and $n_2=N$ into Eq.(\ref{eq:Fock-output}), we find the SU(2) coherent state
\begin{equation}
\label{eq:SU(2)state}
\hat{B}|0,N\rangle = \sum_{k=0}^{N}c_k^N|k,N-k\rangle
\end{equation}
where
\begin{equation}
\label{eq:SU(2)-coefficient}
c_k^N=\pmatrix{
N\cr
k\cr
}^{1/2} r^k t^{N-k}\mbox{e}^{ik\phi}.
\end{equation}
The von Neumann entropy for the reduced density operator
$\hat{\rho}$ is
$E(\hat{\rho}_a)=\sum_{k=0}^{N}|c_k^N|^2\ln|c_k^N|^2$. In Fig. 2,
the von Neumann entropy for $N=10$ is plotted, which shows that
the measure of entanglement is a convex function with its maximum
for a 50:50 beam splitter, {\em i.e.} $r=t=1/2$. In
particular, when $N=1$ the output state is
$1/\sqrt{2}(|0,1\rangle+|1,0\rangle)$ for a 50:50 beam splitter
\cite{Tan,Loudon}.

\subsection{Input fields of same number of photons}
In Fig. 2, it is interesting to note that, for a 50:50 beam
splitter, the entanglement shows a dip when $n_1=n_2$.
When the two input Fock states have the same number of photons,
{\em i.e.}, $n_1=n_2=n$, the output state is
\begin{eqnarray}
\label{eq:sameinput}
\hat{B}&&|n,n\rangle=\sum_{m=0}^{n}\mbox{e}^{-i(n-2m)\phi}\left({1\over 2}\right)^n\sum_{k=0}^{n}(-1)^{n-k}
\pmatrix{
n\cr
k\cr}\nonumber \\
&\times&
\pmatrix{
n\cr
2m-k\cr
}
\frac{\sqrt{2m!(2n-2m)!}}{n!}|2m,2n-2m\rangle
\end{eqnarray}
for a 50:50 beam splitter. This shows that the possibility of
having odd numbers of photons is zero \cite{Lai}.   This is an
extension of the well-known result of $\hat{B}|1,1\rangle={1 \over
\sqrt{2}}(|0,2\rangle+\mbox{e}^{i\phi}|2,0\rangle)$
\cite{Zeilinger}. Output state $|1,1\rangle$ may result from
transmission or reflection of both the photons.  The two cases
destructively interfere to remove the $|1,1\rangle$ state in the
output state. In fact, the output state is the maximally entangled
state in the Hilbert space composed of $|0\rangle$ and
$|2\rangle$.  We now see why entanglement does not maximize when the same
number of photons are injected to a 50:50 beam splitter. This is
due to the fact that odd-number states destructively interfere
and do not appear in the output state.  It is also true that the
output state can be considered in ($n+1$)-dimensional Hilbert
space composed of $|0\rangle,
|2\rangle,\cdot\cdot\cdot,|2n\rangle$ instead of
($2n+1$)-dimensional space.

With the use of a beam splitter, there are two ways to generate
entangled states in ($n+1$)-dimensional Hilbert space. One way is
to put total of $n$ photons into a beam splitter and the other
way is to put $n$ photons into each input port of a 50:50 beam
splitter. By comparing the von Neumann entropies for the both
cases, we find that the latter case of using a 50:50 beam splitter
does not bring about the best entanglement.

\section{Squeezed state inputs}
Generating Gaussian states, in particular, coherent states and
squeezed states has become a standard experimental technique.
When two coherent states are incident on a beam splitter, the
output is given by
\begin{equation}
\label{eq:two-coherent}
\hat{B}\hat{D}_a(\alpha)\hat{D}_b(\beta)|0,0\rangle=\hat{D}_a(t\alpha+r\mbox{e}^{i\phi}\beta)
\hat{D}_b(t\beta-r\mbox{e}^{-i\phi}\alpha)|0,0\rangle
\end{equation}
where $\hat{D}(\alpha)=\exp(\alpha\hat{a}^\dag-\alpha^*\hat{a})$
is the displacement operator \cite{Glauber}. The output state
(\ref{eq:two-coherent}) is clearly not entangled.  It is further
found that {\em displacing the input fields does not increase
entanglement of the output fields} because the impact of
displacing the input fields can always be canceled by local
unitary operations on the output fields.

When the two input fields are squeezed, the output state from a
beam splitter is
\begin{equation}
\label{eq:squeezed}
\hat{B}\hat{S}_a(\zeta_1)\hat{S}_b(\zeta_2)|0,0\rangle
\end{equation}
where the squeezing operator \cite{LoudonKnight}
\begin{equation}
\label{eq:squeezing-operator}
\hat{S}(\zeta)=\exp\left({1\over 2}\zeta^*\hat{a}^2-{1\over 2}\zeta\hat{a}^{\dag 2}\right)
\end{equation}
with the complex squeezing parameter $\zeta=s\exp(i\varphi)$.  The phase $\varphi$ of the
squeezing parameter determines the direction of squeezing.  Using the rotation operator
$\hat{R}(\vartheta)=\exp(i\vartheta\hat{a}^\dag\hat{a})$  the following can be written
\begin{eqnarray}
\label{rotation}
\hat{B}(\theta,\phi)\hat{S}(\zeta) &=&\hat{B}(\theta,\phi)\hat{R}(\varphi/2)\hat{S}(s)
\hat{R}^\dag(\varphi/2)\nonumber \\
&=& \hat{R}(\varphi/2)\hat{B}(\theta,\phi-\varphi/2)\hat{S}(s)\hat{R}^\dag(\varphi/2),
\end{eqnarray}
where, in order to specify the parameters $\theta,~\phi$ of  the
beam splitter operator, the beam splitter operator has been
denoted by $\hat{B}(\theta,\phi)$. The first rotation operator in
the last line of Eq.~(\ref{rotation}) is canceled by local
operation and the last rotation operator does not change the
state when it applies to the vacuum.  Now, we have found that the
relative phase $\phi$ between the amplitude reflection and
transmission coefficients gives the effect of the rotation of the
squeezing angle for the input fields. Without losing generality,
we take the input squeezing parameter to be real while keeping
$\phi$ variable.

The von Neumann entropy $E(\hat{\rho}_a)$ of the output state
(\ref{eq:squeezed}) is plotted in Fig.~3 against squeezing
parameter $s_2$ and reflection coefficient  for $s_1=0.5$.
The relative phase $\phi=0$ in Fig.~3(a) and $\pi/2$ in Fig.~3(b).
We find that the entanglement of the output state depends on the
degrees of squeezing for input fields and the reflection
coefficient.  We also note that the relative phase $\phi$ hence
relative angle of squeezing for input fields plays an important
role. For a 50:50 beam splitter, the entanglement of the output
state is minimized when $\phi=0$, while it is maximized when
$\phi=\pi/2$. In other words, for $\phi=0$, the entanglement of
the output state is maximized if the two input fields are
squeezed along the conjugate quadratures in phase space. To
analyze the output state (\ref{eq:squeezed}) further, consider
the following relation for a 50:50 beam splitter of
$\phi=\ell\pi/2$ ($\ell=0,1,2,\cdot\cdot\cdot$).  In this case,
the output state (\ref{eq:squeezed}) can be written as
\begin{eqnarray}
\label{simple-squeezing}
&&\hat{B}(\pi/4,\phi)\hat{S}_a(s_1)\hat{S}_b(s_2)|0,0\rangle \nonumber \\
&&=\hat{S}_a\left({1\over 2}(s_1+s_2\mbox{e}^{2i\phi})\right)\hat{S}_b\left({1\over 2}(s_1\mbox{e}^{-2i\phi}+s_2)\right)
\nonumber \\
&&\times\hat{S}_{ab}\left({1\over 2}(s_1\mbox{e}^{i\phi}-s_2\mbox{e}^{-i\phi})\right)|0,0\rangle
\end{eqnarray}
where
$\hat{S}_{ab}(\zeta)=\exp(-\zeta\hat{a}\hat{b}+\zeta^*\hat{a}^\dag\hat{b}^\dag)$
is the two-mode squeezing operator. The single-mode squeezing
operators $\hat{S}_a$ and $\hat{S}_b$ in the right-hand side of
Eq.(\ref{simple-squeezing}) do not contribute toward entanglement
of the output state because they can be canceled by local unitary
operations.  Thus only the two-mode squeezing operator
$\hat{S}_{ab}$ determines the entanglement of the output state as
only it represents a joint action on both pairs of the bipartite
system. For a given squeezing, $s_1$ and $s_2$, when
$\phi=\pi/2$, the output state is maximally entangled.  When
$\phi=0$, entanglement is minimized.  In fact, if $s_1=s_2$ we
completely lose entanglement for $\phi=0$.  We notice that {\it a
two-mode squeezed state is produced from a single-mode squeezed
state by an action of a beam splitter and local unitary
operations}. In contrast to the case of the Fock-state input, the
relative phase between reflection and transmission plays an
important role for the case of squeezed input fields.

So far, we have studied only pure input states. From what we have
learned we can conclude that the nonclassical behavior of the
input fields is a necessary condition for the output fields to be
entangled. Specifically, the only pure state which does not
possess nonclassical properties is a coherent state (Its
$P$-function is positive well-defined. See the discussion in the
next section.). As it is well known coherent inputs never become
entangled in the beam splitter, that is the output can always be
written in the factorized form. On the other hand, as we have
shown above, nonclassicality of the inputs is not a sufficient
condition for the entanglement.

\section{Gaussian mixed state input}
When the input fields are mixed, the output fields from a beam
splitter are also mixed.  A general mixed continuous-variable
state is not easy to deal with because of its complicated
nature.  However, for a Gaussian two-mode state, the separability
condition has been studied extensively \cite{Simon,Duan,Kim}.

The separability of a Gaussian state is discussed with
quasi-probability functions and their characteristic functions in
phase space. There are a group of quasi-probability functions
including the Wigner function, the Husimi $Q$ function and the
$P$-function \cite{Cahill}.  In particular, the $P$-function can
be used as the measure of the nonclassicality of the given
field.  For example, if a single-mode state is nonclassical its
density operator $\hat{\rho}^{cl}$ can be written as
\begin{equation}
\label{non-classical-state-P} \hat{\rho}^{cl}=\int
P(\alpha)|\alpha\rangle\langle\alpha|d^2\alpha
\end{equation}
where the $P$-function $P(\alpha)$ is positive and well-behaved.

It has been shown that if a two-mode Gaussian state is represented by a positive well-behaved
$P$-function $P(\alpha,\beta)$, the state is separable \cite{Duan,Kim}.  Suppose two classical states
of $P$-functions $P_a(\alpha)$ and $P_b(\beta)$ are incident on a beam splitter.  Using
Eq.(\ref{non-classical-state-P}), the density operator for the output state is written as
\begin{eqnarray}
\label{two-mode-P} &&\hat{B}\int
P_a(\alpha)P_b(\beta)|\alpha\rangle_a\langle\alpha|\otimes|\beta\rangle_a\langle\beta|
d^2\alpha d^2\beta\hat{B}^\dag
\nonumber \\
&=&\int P_a(\alpha)P_b(\beta)|t\alpha+r\mbox{e}^{i\phi}\beta\rangle_a\langle t\alpha+r\mbox{e}^{i\phi}\beta|
\nonumber \\
&&~~~~~~~~~~\otimes
|-r\mbox{e}^{-i\phi}\alpha+t\beta\rangle_b\langle-r\mbox{e}^{-i\phi}\alpha+t\beta|d^2\alpha d^2\beta
\nonumber \\
&=&\int P_a(t\gamma-r\mbox{e}^{i\phi}\delta)P_b(r\mbox{e}^{-i\phi}\gamma+t\delta)|\gamma\rangle_a\langle\gamma|\otimes
|\delta\rangle_a\langle\delta|
d^2\gamma d^2\delta.
\end{eqnarray}
Here $P_a(t\gamma-r\mbox{e}^{i\phi}\delta)P_b(r\mbox{e}^{-i\phi}\gamma+t\delta)$ is the two-mode
$P$-function for the output state. Because $P_a(\alpha)$ and
$P_b(\beta)$ are positive well-defined under the assumption of
classical input fields,
$P_a(t\gamma-r\mbox{e}^{i\phi}\delta)P_b(r\mbox{e}^{i\phi}\gamma+t\delta)$ is also positive
well-defined. We have proved {\em a sufficient condition} for
separability of the output state from a beam splitter: when two
{\em classical} Gaussian input fields are incident on a beam
splitter, the output state is always separable. It follows that
for creating a Gaussian entangled state with a help of a beam
splitter, it is necessary that the input exhibits nonclassical
behavior.

We have already seen that two nonclassical input fields do not necessarily bring about entanglement
in the out state as two squeezed state inputs may not entangled in the beam splitter.
We investigate the entanglement of the output state when two Gaussian mixed states are incident on
a beam splitter.

The necessary and sufficient criterion for the separability of a Gaussian mixed state has been
studied using the Weyl characteristic function $C^{(w)}(\zeta,\eta)$ \cite{Simon,Duan,Kim}.
For a two-mode Gaussian state of density operator $\hat{\rho}_{ab}$,
the Weyl characteristic function \cite{Barnett},
$C^{(w)}(\zeta,\eta)\equiv\mbox{Tr}\hat{\rho}_{ab}\hat{D}_a(\zeta)\hat{D}_b(\eta)$, can be written as
\begin{equation}
\label{Weyl}
C^{(w)}(\zeta,\eta)=\exp\left[-{1\over 2}(\zeta_i,\zeta_r,\eta_i,\eta_r)M(\zeta_i,\zeta_r,\eta_i,\eta_r)^T\right]
\end{equation}
where $M$ is a $4\times 4$ matrix which completely determines the
statistical properties of the Gaussian state. Duan {\em et al.}
\cite{Duan} found that after some local operations, it is possible
to transform the state into another that is represented by the
matrix,
\begin{equation}
\label{modified-matrix}
M'=\pmatrix{
b_1&0&c_1&0\cr
0&b_2&0&c_2\cr
c_1&0&d_1&0\cr
0&c_2&0&d_2
}
\end{equation}
where parameters $b_i$, $d_i$ and $c_i$ satisfy
\begin{eqnarray}
\label{parameter1}
\frac{b_1-1}{d_1-1} &=&\frac{b_2-1}{d_2-1}
\\
\label{parameter2}
|c_1|-|c_2| &=&\sqrt{(b_1-1)(d_1-1)}
\nonumber \\
&&-\sqrt{(b_2-1)(d_2-1)}.
\end{eqnarray}
Note that parameters $c_{1,2}$ determine the correlation between two modes.
The necessary and sufficient criterion for separability reads then
\begin{equation}
\label{necessary-sufficient}
\langle(\Delta \hat{u})^2\rangle+\langle(\Delta \hat{v})^2\rangle\geq q_o^2+{1\over q_o^2}
\end{equation}
where $q_o^2=\sqrt{(d_i-1)/(b_i-1)}$ and two operators $\hat u$ and $\hat v$ are defined as
\begin{eqnarray}
\label{epr-operators}
\hat{u}={q_o \over \sqrt{2}}(\hat{a}+\hat{a}^\dag)-\frac{c_1}{|c_1|}\frac{1}{\sqrt{2}q_o}(\hat{b}^\dag+\hat{b})
\nonumber \\
\hat{v}={iq_o \over \sqrt{2}}(\hat{a}^\dag-\hat{a})
-\frac{c_2}{|c_2|}\frac{i}{\sqrt{2}q_o}(\hat{b}^\dag-\hat{b}).
\end{eqnarray}

When two mixed states of density operators $\hat{\rho}_a$ and $\hat{\rho}_b$ are input to a beam splitter,
the density operator for the two-mode output field is $\hat{\rho}_{out}=\hat{B}\hat{\rho}\hat{B}^\dag$.
The Weyl characteristic function for the output field is
\begin{equation}
\label{Weyl-definition}
C^{(w)}_{out}(\zeta,\eta)= C^{(w)}_a(t\zeta+r\mbox{e}^{i\phi}\eta)C^{(w)}_b(-r\mbox{e}^{-i\phi}\zeta+t\eta)
\end{equation}
which is obtained using the relation $\hat{B}^\dag\hat{D}_a(\zeta)\hat{D}_b(\eta)\hat{B}=
\hat{D}_a(t\zeta+r\mbox{e}^{i\phi}\eta)\hat{D}_b(-r\mbox{e}^{-i\phi}\zeta+t\eta)$.

\subsection{Squeezed thermal state inputs}
Consider two thermal states of the same average photon number
$\bar n$. The density operator for the thermal field is
\cite{Loudon}
\begin{equation}
\label{termal-density}
\hat{\rho}_{th}=\sum_n\frac{(\bar n)^n}{(1+\bar n)^{1+n}}|n\rangle\langle n|.
\end{equation}
Suppose the thermal fields are respectively squeezed before they
are mixed at a beam splitter. From the earlier section, we know
that two squeezed vacua result in maximum entanglement for the
output field when $\phi=\pi/2$. We thus restrict our discussion
to the case $\phi=\pi/2$ for the study of two squeezed thermal state
inputs. We also assume that the incident fields are equally
squeezed.

The squeezed thermal field $\hat{S}(s)\hat{\rho}_{th}\hat{S}^\dag(s)$ is represented by the following characteristic function:
\begin{equation}
\label{squeezed-th-char}
C^{(w)}(\zeta)=\exp\left[-{1 \over 2}(2\bar n+1)\mbox{e}^{2s}\zeta_r^2-{1 \over 2}(2\bar n+1)\mbox{e}^{-2s}\zeta_i^2
\right].
\end{equation}
The squeezed thermal state is said to be nonclassical when one of the quadrature variables has its variance smaller than
the vacuum limit; the squeezed thermal state of (\ref{squeezed-th-char}) is nonclassical when \cite{Kim+Knight}
\begin{equation}
\label{non-classical}
(2\bar n+1)\mbox{e}^{-2s}-1 < 0.
\end{equation}
Throughout the paper $s>0$ is assumed without loss of generality.

For the maximum entanglement of the squeezed input, let us consider a 50:50 beam splitter.  Substituting $C^{(w)}_{a,b}$ of
(\ref{squeezed-th-char}) into Eq.(\ref{Weyl-definition}), the matrix elements in Eq.(\ref{Weyl}) are found:
\begin{eqnarray}
\label{sq-thermal-elements}
b_1=b_2&=&d_1=d_2={ 1\over 2}(2\bar{n}+1)(\mbox{e}^{2s}+\mbox{e}^{-2s})
\nonumber \\
c_1&=&{1 \over 2} (2\bar n+1)(\mbox{e}^{-2s}-\mbox{e}^{2s})
\nonumber \\
c_2&=&{1 \over 2} (2\bar n+1)(\mbox{e}^{2s}-\mbox{e}^{-2s}).
\end{eqnarray}
The separability condition (\ref{necessary-sufficient}) in this case reads that the output state is separable
when $b_1-1\geq |c_1|$.   Substituting $b_1$ and $c_1$ in (\ref{sq-thermal-elements}), it is found that
the output state is
separable when $(2\bar n+1)\mbox{e}^{-2s}-1 \geq 0$.  With help of  Eq.(\ref{non-classical}), we write
that the output state is
entangled when the squeezed thermal input fields becomes nonclassical.

\subsection{Squeezed thermal and vacuum input states}
Suppose a squeezed thermal state is incident on one input port
and vacuum is incident on the other input port.  As was done
earlier, we assume $\phi=\pi/2$ for the beam splitter.  In this
subsection we release the condition of the 50:50 beam splitter
hence the output state depends on the reflection coefficient of the
beam splitter. The output state is then represented by the matrix
$M$ with its elements:
\begin{eqnarray}
b_1=r^2(2\bar n+1)\mbox{e}^{-2s}+t^2~&;&~b_2=r^2(2\bar n+1)\mbox{e}^{2s}+t^2
\nonumber \\
d_1=t^2(2\bar n+1)\mbox{e}^{-2s}+r^2~&;&~d_2=t^2(2\bar n+1)\mbox{e}^{2s}+r^2
\nonumber \\
c_1=tr[(2\bar n+1)\mbox{e}^{-2s}-1]~&;&~c_2=tr[(2\bar n+1)\mbox{e}^{2s}-1].
\label{elements-sq-vac}
\end{eqnarray}
The separability criterion (\ref{necessary-sufficient}) takes different forms depending on the positivity of $b_1-1$ and
$d_1-1$ due to the definition of $q_o$.   When $b_1-1\geq 0$ and $d_1-1\geq 0$, the separability criterion becomes
\begin{equation}
\label{separability-positive}
\sqrt{(b_1-1)(d_1-1)}+\sqrt{(b_1-1)(d_1-1)}\geq |c_1|+|c_2|.
\end{equation}
Otherwise the separability criterion is
\begin{equation}
\label{separability-negative}
-\sqrt{(b_1-1)(d_1-1)}+\sqrt{(b_1-1)(d_1-1)}\geq |c_1|+|c_2|
\end{equation}
With the use of $b_1$ and $d_1$ in (\ref{elements-sq-vac}), we find that both conditions $b_1-1\geq 0$ and $d_1-1\geq 0$ imply
$(2\bar n+1)\mbox{e}^{-2s}-1 \geq 0$.  In this case, the inequality (\ref{separability-positive}) is always satisfied
and the output state is separable.  However, when $(2\bar n+1)\mbox{e}^{-2s}-1 < 0$, the separability criterion
(\ref{separability-negative}) is never satisfied and the output state is entangled.  Here, we confirm our earlier finding that
the nonclassicality of the input state provides the entanglement criterion for the output state.
When a squeezed thermal state and vacuum are incident on a beam splitter, the output state is entangled
only if the squeezed thermal state is nonclassical.

\subsection{Squeezed vacuum and thermal input states}
So far, we found that nonclassicality of the incident field plays an important role in the entanglement of the output field.
Let us suppose that an input field is a squeezed vacuum and the other input field is a thermal state.  Differently from the earlier
cases in this section, one of the input states is always nonclassical while the other is always classical.  Substituting
the characteristic functions for the thermal state and squeezed state into Eq.(\ref{Weyl-definition}), the characteristic
function for the output field is represented by (\ref{Weyl}) with matrix $M$ in the form (\ref{modified-matrix}) and the matrix elements are
\begin{eqnarray}
b_1=(2\bar n+1)r^2+\mbox{e}^{-2s}t^2~&;&~b_2=(2\bar n+1)r^2+\mbox{e}^{2s}t^2
\nonumber \\
d_1=(2\bar n+1)t^2+\mbox{e}^{-2s}r^2~&;&~d_2=(2\bar n+1)t^2+\mbox{e}^{2s}r^2
\nonumber \\
c_1=tr(2\bar n+1-\mbox{e}^{-2s})~&;&~c_2=tr(2\bar n+1-\mbox{e}^{2s}).
\label{element-th-sq}
\end{eqnarray}
These elements do not satisfy conditions (\ref{parameter1}) and (\ref{parameter2}).  In order to use the
separability criterion (\ref{necessary-sufficient}), the output state has to be locally transformed.

Suppose the output fields are squeezed locally.  Assuming equal degree of squeezing, $s$, for each mode,
the transformed state is represented by $\hat{\rho}_{trans}
=\hat{S}_a\hat{S}_b\hat{\rho}_{out}\hat{S}_a^\dag\hat{S}_b^\dag$.  We use the identity $\hat{S}^\dag (s)\hat{D}(\alpha)\hat{S}(s)
=\hat{D}(\alpha_r\mbox{e}^{s}+i\alpha_i\mbox{e}^{-s})$, where sub-indeces $_r$ and $_i$ respectively denote real and imaginary
parts, and definition (\ref{Weyl}), to find the Weyl characteristic function for the transformed state:
\begin{equation}
C_{trans}^{(w)}(\zeta,\eta)=C_{out}^{(w)}\left(\zeta_r\mbox{e}^s+i\zeta_i\mbox{e}^{-s},\eta_r\mbox{e}^s+i\eta_i\mbox{e}^{-s}\right)
\label{trans-character}
\end{equation}
where $C_{out}^{(w)}$ is the characteristic function for the output state.  After a little algebra, we find that the matrix
elements representing $C_{trans}^{(w)}(\zeta,\eta)$ is the same as those in (\ref{elements-sq-vac}) for the output state
from a beam splitter when the squeezed thermal and vacuum are input fields but with squeezing factor
$-s$.  The separability criterion $(2\bar n+1)\mbox{e}^{-2s}-1 < 0$, thus, applies for the output state
when the two input fields are the squeezed vacuum and thermal field. The separability criterion coincides with the nonclassicality
condition for the output field of mode $c$ in Fig. 1.

\section{Remarks}
We have considered the nature of the entanglement of output
fields from a beam splitter for pure state inputs and for mixed
Gaussian state inputs. In the case of pure states we have found
that  for Fock state inputs, the
beam splitter is a tool to produce a ($N+1$)-dimensional entangled
state, where $N$ is the total excitation of the input fields. For
squeezed vacuum inputs, the entanglement of output fields depends
on many factors including the relative angle of squeezing between
two input fields.  When the relative angle is appropriately
chosen, the entanglement of the output state is maximized for a
50:50 beam splitter.  From these results it directly follows
that nonclassicality of input pure states is a necessary
condition for having entangled states at the output of the
beam splitter.

In the case of mixed states the analysis is more complicated
since there does not exist a necessary and sufficient condition
for inseparability of arbitrary infinite-dimensional bi-partite
systems. Since the  condition exists for Gaussian states, we have
concentrated our attention on these states. We have proved a
sufficient condition for
the output state of a beam splitter to be separable (that is they are
not entangled): if both the Gaussian input fields are classical,
it is not possible to create entanglement in the output of the
beam splitter. From here it automatically follows that
nonclassicality is a necessary condition for the entanglement.

These observations make us conjecture that
nonclassicality of at least one of the input fields
is a necessary condition for the output to be entangled.
That is the nonclassicality of individual inputs
can be traded for quantum entanglement of the output
of the beam splitter.

\acknowledgements

This work was supported by the UK Engineering and Physical
Sciences Research Council (EPSRC), the BK21 Grant of the Korea
Ministry of Education and the European Union project EQUIP under
contract IST-1999-11053. WM thanks Dr. J. Lee for discussion and Korean Ministry
of Science and Technology through the Creative Research
Initiatives program for financial support under contract No. 00-C-CT-01-C-35.

\begin{figure}
  \begin{center}
\centerline{\epsfig{width=7.0cm,file=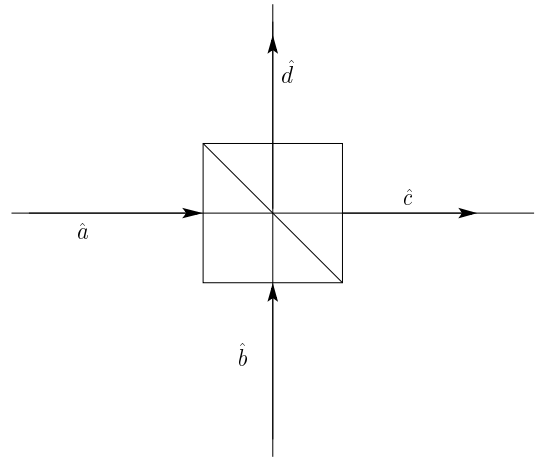}}
    \caption{
Configuration of beam splitter operation.}
  \label{fig:beam-splitter}
  \end{center}
\end{figure}

\begin{figure}
  \begin{center}
\centerline{\scalebox{.3}{
\includegraphics{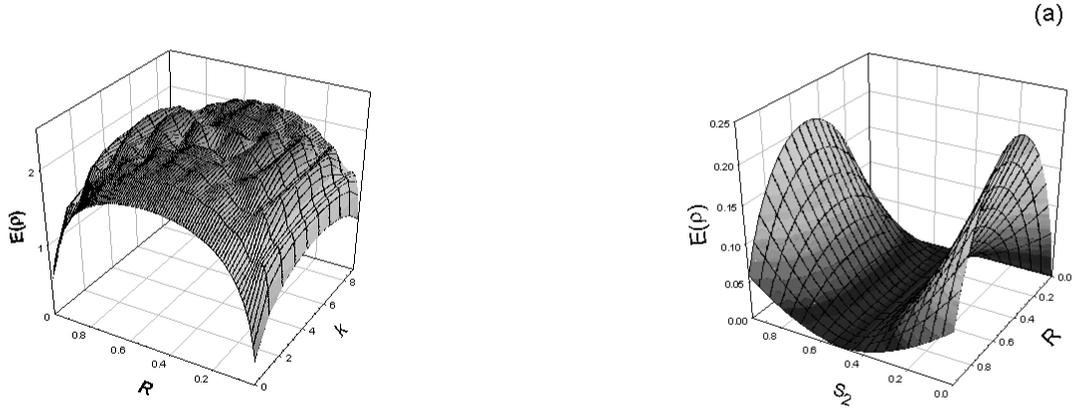}}}
    \caption{
The measure of entanglement $E(\hat{\rho})$ is plotted using the
von Neumann entropy for the reduced density operator of the
beam-splitter output field.  The Fock-state input fields
$|k,N-k\rangle$ have total photon number $N=10$.  $R\equiv r^2$.}
  \label{fig:entanglement-Fock}
  \end{center}
\end{figure}

\begin{figure}
  \begin{center}
\centerline{\scalebox{.3}{
\includegraphics{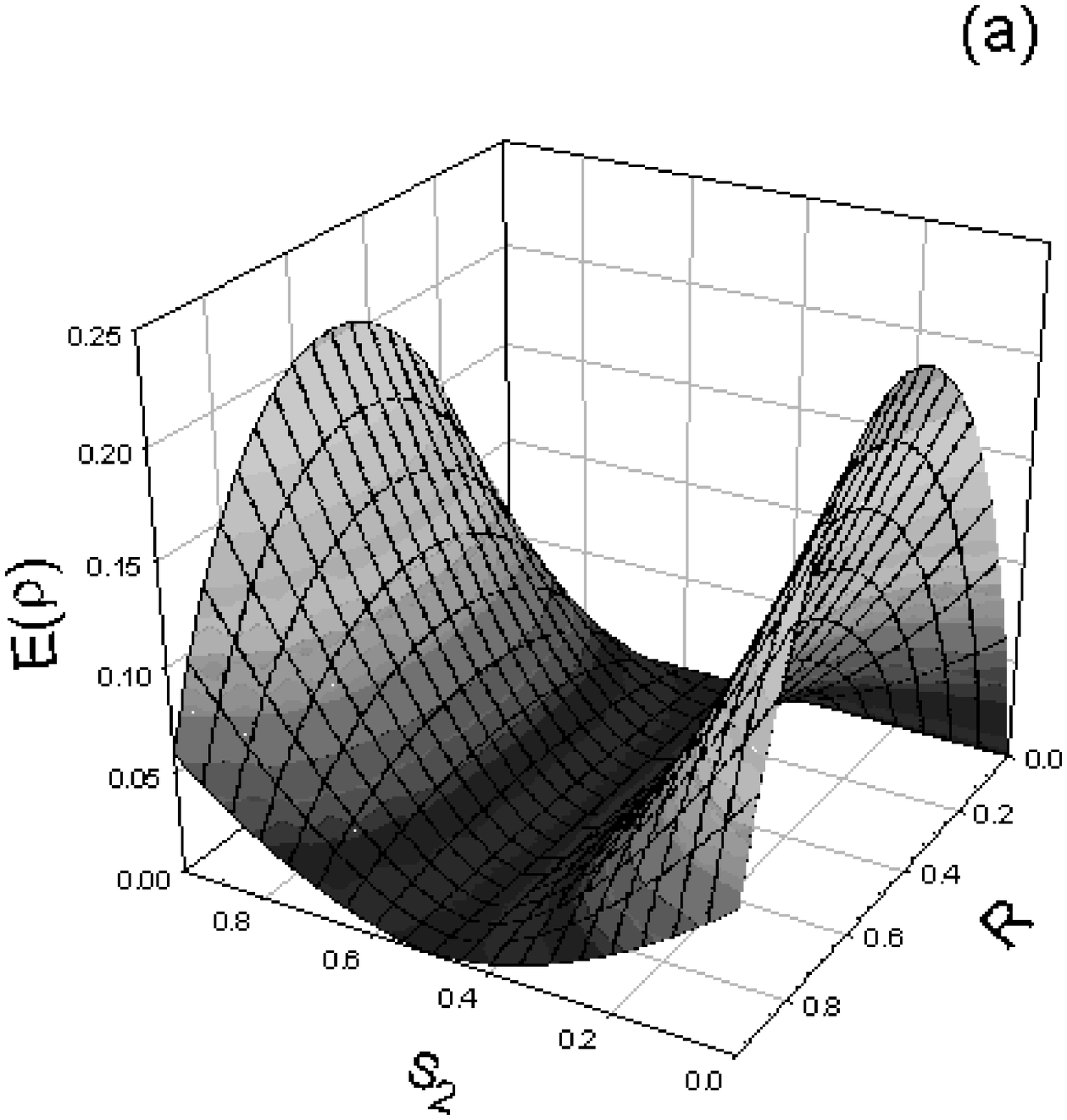}}}
\centerline{\scalebox{.3}{
\includegraphics{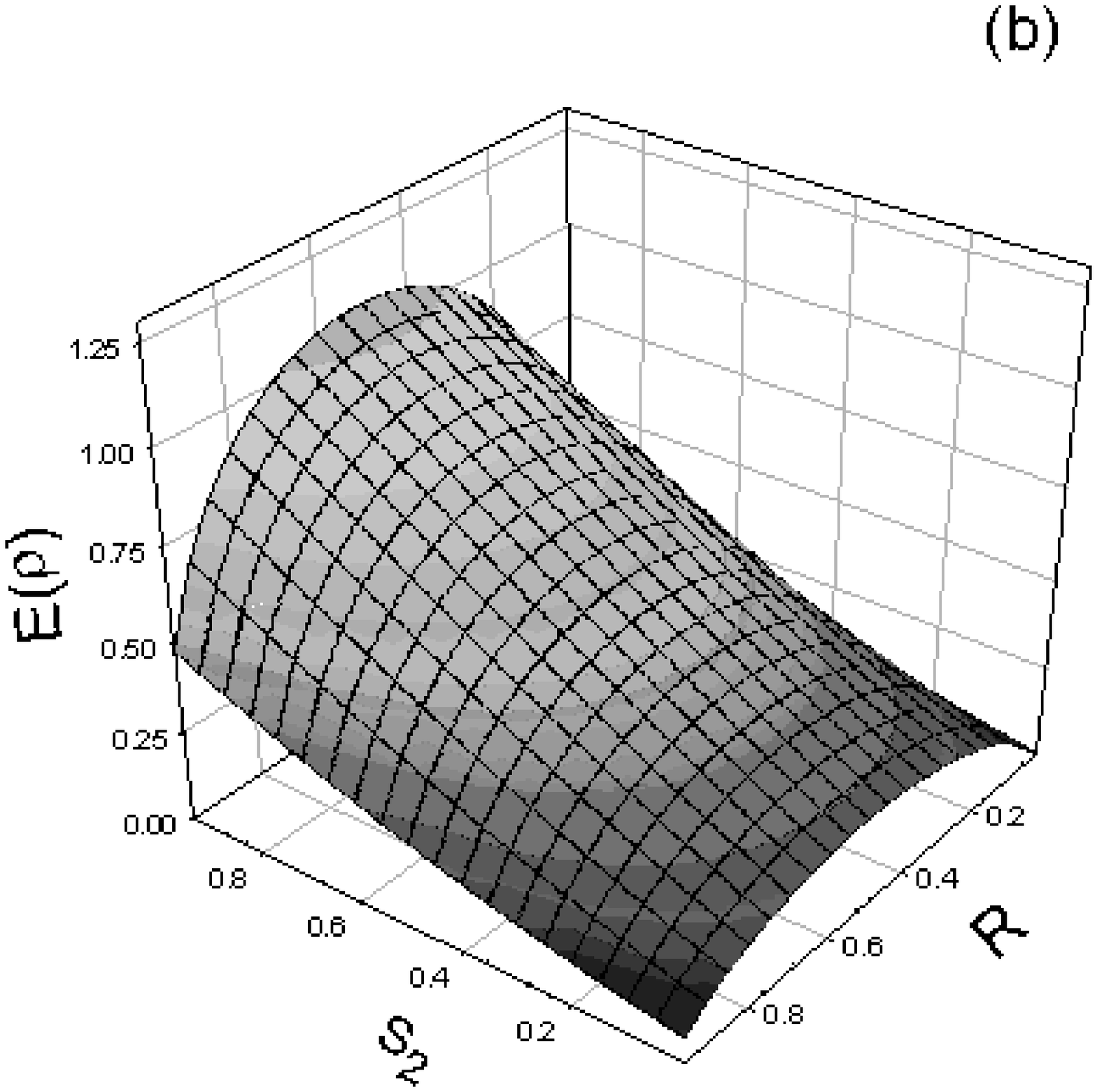}}}
    \caption{
The measure of entanglement $E(\hat{\rho})$ for the beam-splitter
output field is plotted using the von Neumann entropy for the
reduced density operator of the output field.  The squeezing
parameter for one squeezed input is fixed to $s_1=0.5$ while the
squeezing parameter for the other squeezed state is varied from
$s_2=0$ to 1. The transmittivity $R$. The beam splitter gives
phase difference $\phi=0$ (a) and $\phi=\pi/2$ (b) between the
reflected and transmitted fields. $R\equiv r^2$.}
  \label{fig:entanglement-squeezed}
  \end{center}
\end{figure}

\end{document}